\documentclass[aps,prb]{revtex4-1} 

\usepackage{amsmath}  
\usepackage{amsfonts} 
\usepackage{graphicx} 

\begin{document}


\title{Phase-Sensitive Detection in the undergraduate lab using a low-cost microcontroller}


\author{K.\ D.\ Schultz}
\email[]{schultzk@harwick.edu}
\homepage[]{www.HartwickChaosLab.github.io}

\affiliation{Hartwick College, Oneonta NY 13820 USA}


\date{\today}

\begin{abstract}
Phase-sensitive detection (PSD) is an important experimental technique that allows signals to be extracted from noisy data. PSD is also used in modulation spectroscopy and is used in the stabilization of optical sources. Commercial lock-in amplifiers that use PSD are often expensive and host a bewildering array of controls that may intimidate a novice user. Low-cost microcontrollers such as the Arduino family of devices seem like a good match for learning about PSD; however, making a self-contained device (reference signal, voltage input, mixing, filtering, and display) is difficult, but in the end the project teaches students ``tricks'' to turn the Arduino into a true scientific instrument.
\end{abstract}


\maketitle

\section{Introduction\label{sec:intro}}
Lock-in amplification and phase-sensitive detection are important techniques in experimental physics, see~\cite{bib:Yang:2010wf,bib:Libbrecht:2003cf,bib:Wolfson:1991wu} for pedagogical uses of these techniques. Commerical devices are expensive and can be intimidating for new users. Building a home-made instrument can be instructive~\cite{bib:HandH}; however, doing so requires advanced electronics skills that a student may not already have, putting the emphasis on the electronics and not the method. Conversely, it is possible to perform the mixing and filtering on a computer using a computer's sound card or other low-cost data acquisition devices to handle the input and output, see for instance~\cite{bib:Gonzalez:2007td}, but in the author's experience computer driver issues often get in the way of relatively easy implementation. 

In this paper I describe a PSD that uses the popular Arduino microcontroller and the Processing programming environment~\cite{bib:code}. The major design goal was to make the device as self-contained as possible, a task made difficult by the memory and hardware constraints of a typical microcontroller. While these devices are fantastic for controlling robots and basic data-logging, turning them into scientific instruments requires techniques that go beyond what is normally found in the literature. An added benefit is that these techniques can be used to build other types of instrumentation such as function generators and fast DMMs out of a low-cost microcontroller that typically only cost a few tens of dollars.

\section{Arduino \label{sec:arduino}}
Arduino~\cite{bib:arduino,bib:cookbook} is a catch-all term for a family of open-source hardware based on Atmel micro-controllers with a pre-loaded bootloader so that instead of programming the Arduino with the more difficult, but more powerful, AVR instruction set, users can program with a C$^{++}$-like language. Programming and communication can be done via USB or through a set of on-board communication pins. There have been sixteen different Arduino-labeled boards produced to date, each of which has its own unique hardware and memory specifications. In this paper, the Uno R3 is used, simply because it was what was on hand and is one of the cheaper and more basic Arduinos. The heart of the Uno R3 is the Atmel ATmega328 microcontroller~\cite{bib:atmega}. The Uno R3 has 14 digital input/output pins, six of which provide Pulse Width Modulation (PWM) output. It also has six analog inputs, and an on-board $16$MHz oscillator. The ATmega 328 has 32kB of Flash memory to store programs, and 2kB of SRAM for variable storage. The program given in this paper is easily stored in the Flash memory, but the limited SRAM places severe restrictions on the amount of data that can be taken and manipulated on-board the Arduino.

\section{Phase Sensitive Detection\label{sec:psd}}
\begin{figure} 

\includegraphics[]{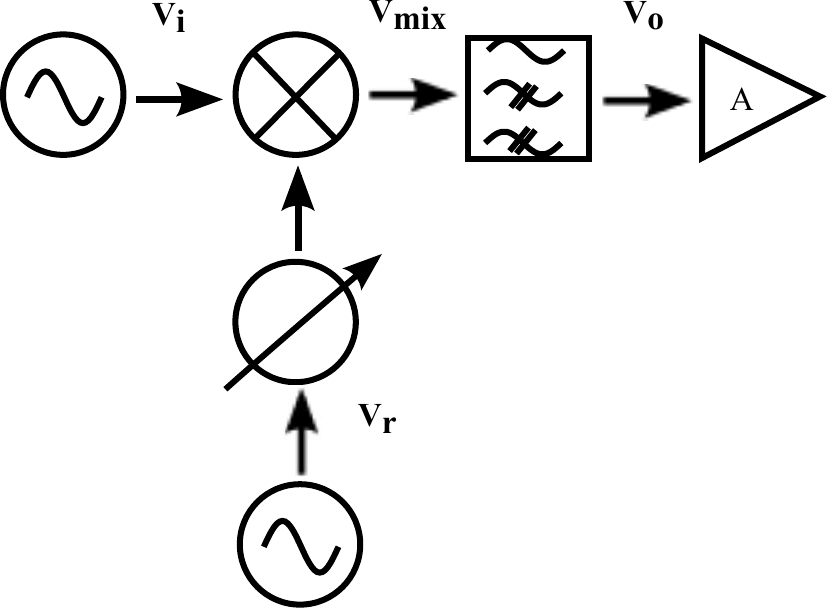}
\caption{Block diagram of a traditional PSD. $V_{i}$ and $V_r$ are the input and reference voltages respectively. \textit{LPF} is a low-pass filter and \textit{A} is an amplifier. A phase-shifter is shown after $V_r$, but this and the amplifier are not used in this project. See the
text for more details.}
\label{fig:block}
\end{figure}

A block diagram for phase-sensitive detection is shown in Fig.~\ref{fig:block}. The input stage of a PSD consists of two signals, the input of interest and the reference signal. Typically, these two signals are oscillating at the same frequency, but in the derivations to follow this will only be enforced at the end. After the signals are acquired, the input signal may be amplified and filtered. After the input stage comes the mixing stage where the input and reference signals are combined. There are a number of ways of doing this, but in this implementation the signals will simply be multiplied. Finally, the output of the mixing stage is heavily filtered by a low-pass filter. Once again this can be done using analog circuitry, but here it will be done mathematically.

\subsection{Generic Phase Sensitive Detection \label{generic}}
Let us assume that the input signal  $V_{i}$  and the reference $V_{r}$ are given by

\begin{align}
\label{eq:inputs}
V_{i}&=V_1+V_1 \sin\left(\omega_{s} t - \phi_{s}\right), \nonumber \\
V_{r}&=V_2 \sin \left(\omega_{r} t - \phi_{r}\right),
\end{align}

A DC offset is included with $V_{i}$ because Arduino, like most microcontrollers, only input or output positive voltage; therefore the Arduino outputs an offset reference voltage. This offset is removed during mixing, so $V_{r}$ is centered on zero in what follows. Upon multiplication of the two signals in Eq.~\ref{eq:inputs}, and making use of a trig identity, the output of the mixing stage $V_{mix}$ is 
\begin{widetext}
\begin{equation}
\label{eq:mix}
V_{\mathrm{mix}}=V_1V_2\sin\left(\omega_2 t- \phi_2\right)+\frac{V_1V_2}{2} \lbrace \cos\left[ \left(\omega_2-\omega_1 \right)t-\left(\phi_2-\phi_1\right) \right]-\cos\left[ \left(\omega_2+\omega_1 \right)t-\left(\phi_2+\phi_1\right) \right]\rbrace
\end{equation}
\end{widetext}

At this stage Eq.~\ref{eq:mix} is nothing more than what we find in heterodyne detection in radio and optical engineering. The power of coherent detection is that small signals get amplified by a larger local oscillator signal is evident in Eq.~\ref{eq:mix}. Upon mixing of the two signals there are contributions to the signal at the original frequencies and at the sum and difference frequencies. This is the heart of PSD; the output filter is set to reject all frequencies other than the difference frequency of the inputs. Furthermore, if $V_{i}$ and $V_{r}$ are made to oscillate at the same frequency before entering the PSD, the output $V_{o}$ of the PSD simply depends on the phase difference between the two signals

\begin{equation}
\label{eq:output}
V_{o}=\frac{V_1V_2}{2}\cos\left(\phi_o-\phi_1\right).
\end{equation}

The final effect of this filtering is to move our output from  $\omega_{1,2}$ to DC. The stronger the filtering, the more noise is rejected and the signal-to-noise increases. However, in effect, the PSD is performing signal averaging, so with each factor of two increase in the signal-to-noise, the collection time increases by a factor of four. The other disadvantage of PSD is that by making the measurements at DC we are placing our signal where $1/f$ noise dominates\cite{bib:Hobbs}. Once again we can see similarities with homodyne detection, and even though we have added no active amplification, the presence of a strong local oscillator can boost a weak experimental signal.

\section{Implementation \label{sec:implementation}}
As stated in Sec.~\ref{sec:intro}, the ultimate goal of this project was to make the project as self-contained as possible. It would need only a computer for display, the Arduino, and as few passive circuit components as possible. These turned out to be fairly severe constraints and required some exploits that are not commonly presented in introductions to microcontrollers. The structure of this section is to discuss the implementation of each of the sub-systems shown in Fig.~\ref{fig:block}. Briefly, and following Fig.~\ref{fig:flowchart}, the Arduino synthesizes a sine wave output from a wavetable. As the output is updated from the wavetable, the input signal is quickly read. Finally after the Arduino has cycled through the complete wavetable, mixing is performed mathematically onboard the Arduino, and the resulting waveforms are sent to the host computer running Processing for filtering and display.

\begin{figure}
\includegraphics{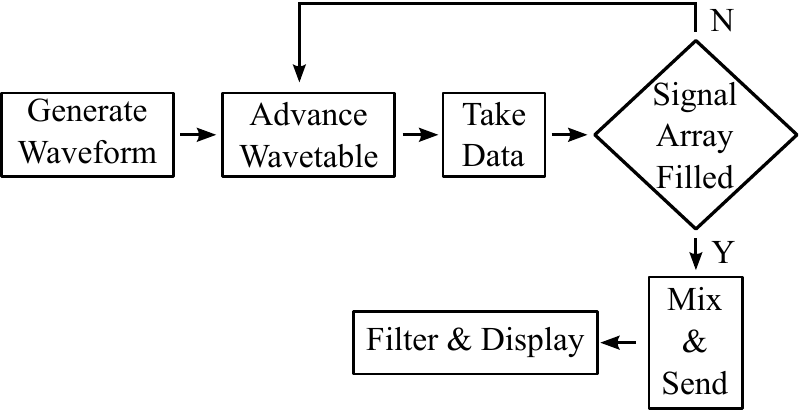}
\caption{All the steps through the mixing of the signals are done on the Arduino. The filtering and displaying are done in a Processing sketch.}
\label{fig:flowchart}
\end{figure}

\subsection{Creating a Reference Signal \label{sec:ref}}
Creating the reference signal is in many ways the most significant constraint on the project. The Arduino does not have a true analog output. The only way to approximate an analog signal is to use PWM with a low pass filter circuit on the output to smooth things out. Additional tricks are needed to generate something resembling a sine wave. 

To generate a reference signal under these limitations a technique sometimes called ``bit-banging''\cite{bib:bitbang} is employed. Bit-banging takes advantage of the timers on the Arduino and PWM. In PWM, the duty-cycle (ratio of `on' to `off' times) of a square wave is modulated. The greater the duty-cycle, the longer the PWM pin on the Arduino is held HIGH. Integrating that square-wave produces a DC voltage that increases with the duty-cycle. The final step is to rapidly change this DC voltage so that the desired waveform, in this case a sine-wave, is synthesized. 

This paper will only briefly sketch the relevant idea. For details, refer to the comments in the source code for this project, the ``bit-banging'' paper~\cite{bib:bitbang}, or the ATmega328 datasheet~\cite{bib:atmega}. This technique relies on two counters on the Arduino. Timer 1, TCNT1, is a 16-bit timer/counter that is configured as an 8-bit counter. TCNT1 runs at the full Arduino clock frequency, $16\mathrm{MHz}$, and produces the PWM output signal. Timer 2, TCNT2, runs slower than TCNT1 by a factor of eight, and is used to step through the pre-generated wavetable. 
\begin{figure}
\includegraphics[angle=90]{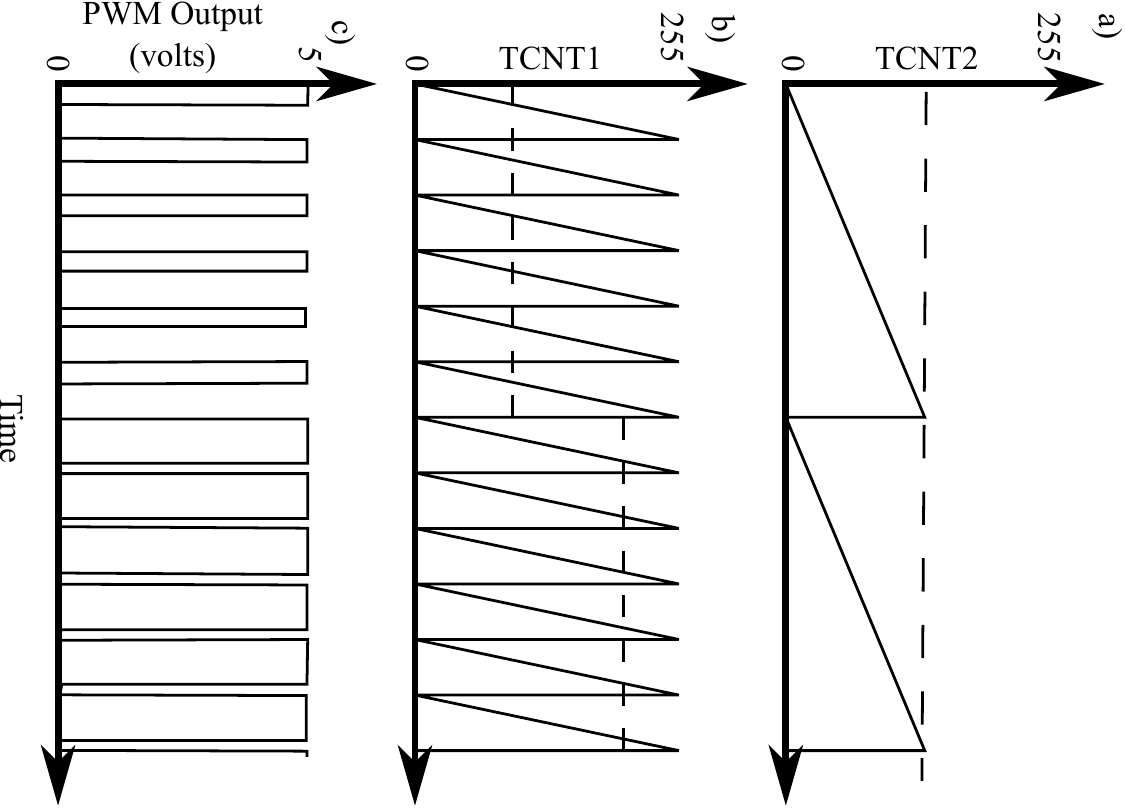}
\label{fig:bitbang}
\caption{The ``bit-banging'' technique requires a slow counter a), a fast counter b), and a pulse-width modulated signal c). The dashed line in a) represents OCR2A, the counter value that calls the necessary interupt to update OCR1A and read in the input voltage for mixing. The dashed line b) represents the changing OCR1A, which sets the duty-cycle for the PWM. The reference signal that modulates the experiment is the output of a low-pass RC-circuit with c) as the input. }
\end{figure}

Output Compare Registers (OCRnA) set a flag when the value of TCNTn matches OCRnA. When the OCRnA flag is set each timer can exhibit different behavior depending on how the Arduino is programmed. When the OCR1AL flag is set, the PWM pin is made to go low, but TCNT1 keeps incrementing until it overflows. At which point TCNT1 starts counting again, the PWM pin goes high again, and OCR1LA is reset. So changing the value of OCR1AL determines the duty-cycle of the PWM signal and consequently the DC voltage that is used to construct $V_{r}$. The pre-programmed wavetable for $V_r$ is what updates OCR1AL. To do this TCNT2, running eight times slower than TCNT1, updates OCR1AL with the next value from the wavetable upon reaching OCR2A. OCR2A is pre-set in the Arduino in the code such that the frequency of $V_r$ is given by:
\begin{align}
\label{eq:f_ref}
f_r= \frac{\textrm{rate of TCNT2}}{\textrm{OCR2A} \times \textrm{wavetable length}}.
\end{align}
When TCNT2 reaches OCR2A, an interrupt is called which does two things, update OCR1AL to the next value of the wavetable and get an input voltage for the ADC. The last step in generating $V_r$ is to place a simple low-pass RC filter on the PWM pin to get a smooth sine-wave.

\subsection{Getting the signal in \label{sec:sig}}

The Arduino ADC is a 10-bit successive approximation circuit connected to an 8-channel multiplexer. The measurements are single-ended and normally referenced to a $1.1 \mathrm{V}$ on-board reference voltage, although it is possible to use an external reference voltage. Because interrupts are used to create an analog output for the reference signal, it is important to ensure that anything that happens during the interrupt is quick, which is why in Fig.~\ref{fig:flowchart}, the data is sent for filtering after the wavetable has been completely cycled through. Serial calls are slow. 

According to ATMEL the ultimate sampling rate for single-ended measurements is limited by the ADC clock speed. The ADC clock speed is derived from the main system clock, and for maximum resolution should be between $50 \mathrm{kHz}$ and $200\mathrm{kHz}$, however, for purposes of this project satisfactory resolution and accuracy are obtained by setting the ADC clock to a speed of $1\mathrm{MHz}$. The successive approximation circuit requires $13$ clock cycles, giving a sampling rate of $77\mathrm{kHz}$~\cite{bib:avr_adc}, which is much faster than this project needs.

\subsection{Phase shifts and mixing \label{sec:phase_shift}}
Since the values of the reference signal are pre loaded into the Arduino memory, changing the phase while data is being taken is difficult. Often, however, the ability to shift the relative phase between $V_i$ and $V_r$ is need to maximize the signal. Additionally, sometimes what is needed in a measurement is the phase shift between the reference and input signals. To meet these needs a cue is taken from dual-phase, lock-in amplifiers\cite{bib:Meade:1982ur} and in addition to our reference waveform a quadrature signal, which is just the reference waveform shifted by $90^\circ$, is generated. Both the in-phase, $I=V_{\textrm{in}}\cos(\omega_rt)$ and quadrature $Q=V_{\textrm{in}}\sin(\omega_r t)$ mixed waveforms are used so that the phase ambiguity is removed by calculating the magnitude signal $R=\sqrt{I^2+Q^2}$ and the phase difference between the input and reference signals is $\tan \phi=Q/I$. The $I$ and $Q$ signals are what are sent from the Arduino to the host computer for filtering and display.

\subsection{Filtering and display \label{sec:processing}}

Once the Arduino finishes running through the waveform, it takes the collected data and sends it via USB to a computer running the Processing IDE~\cite{bib:processing, bib:ProcessingBook}. Processing is a programming language and development environment that was designed to make it easier for the arts community to become software literate. Like the Arduino, it has a vibrant community that has produced numerous tutorials, examples, and books that make learning Processing relatively easy. Processing is also a direct forebear of the Arduino development system and therefore makes it a natural fit with the Arduino side of this project.

In this project the Processing sketch (Processing-talk for program) initiates serial communications with the Arduino and once communications are established, the Arduino sends the $X$ and $Y$ data to the Processing sketch. The Processing sketch applies a recursive, single-pole, low-pass filter\ref{bib:dsp_guide} to the array that is to be displayed ($I,Q,R,\phi$) . In general a recursive filter is given by

\begin{align}
y[n]=a_0 x[n]&+a_1x[n-1]+a_2x[n-2]+\dots\\
&+b_1 y[n-1]+b_2y[n-2]+\dots,
\end{align}

where $y$ is the output of the filter and $x$ is the input data. For this project a single-pole filter was used, so the only coefficients used are given by a single parameter $x_{\textrm{decay}}$

\begin{align}
a_0&=1-x_{\textrm{decay}}\nonumber\\
b_1&=x_{\textrm{decay}}\nonumber\\
x&=e^{-2\pi f_c},
\end{align}

where $f_c$ is the time constant of the filter. Mathematically this filter is identical to a single-pole RC-filter in electronics. The stronger the filtering, the better it is for lock-in detection, since inadequate filtering causes the output to oscillate at the reference frequency. Ref.~\cite{bib:dsp_guide} has algorithms for calculating coefficients for higher-order filters with faster roll-off, single-pole filters are limited to attenuations of $6$dB/decade.

\section{Testing \label{sec:testing}}
\begin{figure}
\label{fig:phase_shift}
\includegraphics[]{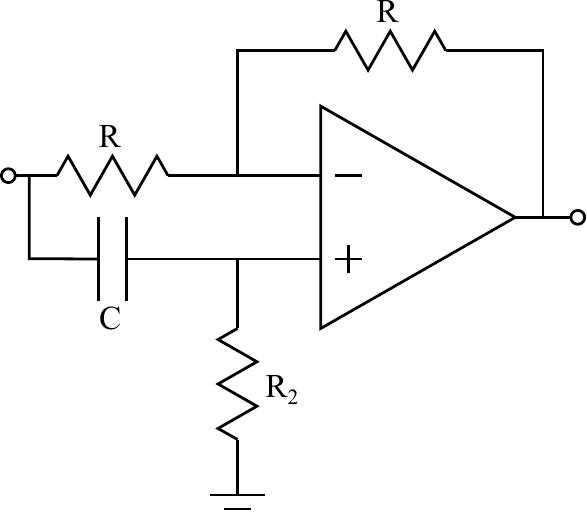}
\caption{Phase shifting circuit for testing the Arduino PSD. The op-amp used was a standard 741. Changing $R_2$ shifts the phase through $180$ degrees.}
\end{figure}

\begin{figure}
\label{fig:results}
\includegraphics[trim=0 0 2in 0, clip, width=3.4in]{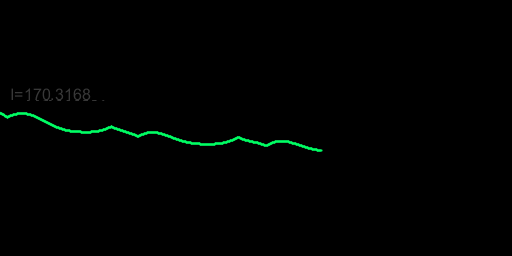}
\caption{The in-phase signal of the Arduino PSD as $R_2$ of the phase-shift test circuit is varied.}
\end{figure}

To test this project a all-pass phase-shifter was built. Fig.~\ref{fig:phase_shift} shows the circuit used to test the Arduino. The voltage gain for this circuit is $1\mathrm{V/V}$, so the amplitude of the output is unchanged with respect to the input voltage. The high-pass RC filter at the non-inverting terminal of the op-amp controls the amount of phase-shift at the output. At $\omega_0=1/RC$ the phase-shift is $90^\circ$ and changes by $90^\circ/\mathrm{decade}$. By varying $R_2$, the $90^\circ$ point shifts and moves our operating frequency along the phase plot. To test the Arduino PSD, the reference signal generated by the Arduino is sent to the input of the phase-shifting circuit. The output of the phase-shifting circuit becomes the input signal to the PSD. Fig.~\ref{fig:results} shows the $I$ signal, as displayed in the Processing sketch, as $R_2$ is varied.

\section{Conclusions \label{sec:conclusions}}
A phase-sensitive detector using an Arduino microcontroller has been described. The main design goal of the project was to make the PSD as self-contained as possible, namely that an external reference signal was not needed and that any phase corrections were unnecessary. Aside from the Arduino and a computer the only other hardware needed are a resistor and a capacitor. Furthermore the techniques used for this project can be used in other projects involving using the Arduino microcontroller as a standalone, low-cost, scientific instrument. The ``bit-banging'' technique is not limited to sine-waves, but can be used to output any synthesized waveform that may be required. More complicated waveforms can be created outside of the Arduino environment and loaded into the EEPROM allowing faster execution and freeing up regular memory.\cite{bib:bitbang}


\begin{acknowledgments}
I would like to thank Prof.\ Lawrence Nienart of Hartwick College for his advice and encouragement throughout this project.
\end{acknowledgments}


\begin{thebibliography}{99}
\bibitem{bib:notation}
In what follows I will use the convention that ``n'' at the end of a name represents a generic counter, and if it is replaced by a number that will designate a specific counter.
\end{thebibliography}
\end{document}